\documentclass[lettersize,journal]{IEEEtran}
\usepackage{amsmath,amsfonts}

\usepackage{algorithmic}
\usepackage{algorithm}

\usepackage{array}
\usepackage[caption=false,font=normalsize,labelfont=sf,textfont=sf]{subfig}
\usepackage{textcomp}
\usepackage{stfloats}
\usepackage{url}
\usepackage{verbatim}
\usepackage{graphicx}
\usepackage{cite}

\usepackage{tablefootnote}
\usepackage{threeparttable}
\usepackage{multirow}
\usepackage{booktabs}
\usepackage{makecell}
\usepackage{colortbl}
\usepackage[table]{xcolor}

\usepackage{latexsym}

\begin{document}

\title{
CoCoEvo: Co-Evolution of Programs and Test Cases to Enhance Code Generation
}

\author{Kefan Li, Yuan Yuan, Hongyue Yu, Tingyu Guo, and Shijie Cao
        % <-this % stops a space
\thanks{Manuscript received 31 January 2025; revised 7 June 2025, accepted 24 July 2025. This work was supported in part by the National Natural Science Foundation of China under Grant 62202023. (Corresponding author: Yuan Yuan.)}% <-this % stops a space
\thanks{Kefan Li is with the School of Computer Science and Engineering, Beihang University, Beijing 100191, China (e-mail: kefanli@buaa.edu.cn).}% <-this % stops a space
\thanks{Yuan Yuan is with the School of Computer Science and Engineering, Beihang University, Beijing 100191, China, also with the Qingdao Research Institute, Beihang University, also with the Hangzhou Innovation Institute, Beihang University, and also with the Zhongguancun Laboratory (e-mail: yuan21@buaa.edu.cn).}% <-this % stops a space
\thanks{Hongyue Yu is with the School of Computer Science and Engineering, Beihang University, Beijing 100191, China (e-mail: Natt1e@buaa.edu.cn).}% <-this % stops a space
\thanks{Tingyu Guo is with the School of Computer Science and Engineering, Beihang University, Beijing 100191, China (e-mail: tingyuguo@buaa.edu.cn).}% <-this % stops a space
\thanks{Shijie Cao is with the School of Computer Science and Engineering, Beihang University, Beijing 100191, China (e-mail: cls1277@buaa.edu.cn).}% <-this % stops a space
}



\maketitle

\begin{abstract}
Large Language Models (LLMs) have shown remarkable performance in automated code generation.
However, existing approaches often rely heavily on pre-defined test cases, which become impractical in scenarios where such cases are unavailable.
While prior works explore filtering techniques between programs and test cases, they overlook the refinement of test cases.
To address this limitation, we introduce CoCoEvo, a novel LLM-based co-evolution framework that simultaneously evolves programs and test cases.
CoCoEvo eliminates the dependency on pre-defined test cases by generating both programs and test cases directly from natural language problem descriptions and function headers.
The framework employs specialized evolutionary operators, including LLM-based crossover and mutation operators for program evolution, along with an additional test case generation operator for test case evolution.
Additionally, we propose optimization strategies such as a crossover rate scheduler to balance exploration and convergence, and a multi-objective optimization method for test case selection.
Experimental results on multiple state-of-the-art LLMs demonstrate that CoCoEvo surpasses existing methods, achieving state-of-the-art performance in automated code generation and testing.
These results underscore the potential of co-evolutionary techniques in advancing the field of automated programming.
\end{abstract}


\section{Introduction}
\IEEEPARstart{I}{n} recent years, Large Language Models (LLMs) have undergone rapid development and have been widely applied in automated software development tasks such as code generation and test case generation. 
Models like GPT-4 \cite{hurst2024gpt}, Llama3 \cite{dubey2024llama}, Qwen2.5 \cite{hui2024qwen2}, and DeepSeek-V3 \cite{liu2024deepseek} have demonstrated exceptional code generation capabilities, significantly advancing the field of automated software development.

Despite their impressive performance, LLMs do not always generate accurate or correct programs.
To enhance the reliability of program generation, test cases are often utilized to evaluate code correctness and provide feedback. 
Existing approaches frequently rely on pre-defined test cases for evaluation. 
For instance, Sampling+Filtering \cite{li2022competition} first generates a batch of candidate codes, evaluates them using pre-defined test cases, and selects the code with the highest score as the final answer. 
Other methods, such as Reflexion \cite{shinn2024reflexion}, Self-Repair \cite{olausson2023self}, INTERVENOR \cite{wang2023intervenor}, LDB \cite{Zhong0S24ldb}, and PairCoder \cite{zhang2024pair}, utilize execution feedback on pre-defined test cases to refine or repair generated code. 
These methods generally assume that pre-defined test cases are accurate and can be fully trusted.

However, pre-defined test cases are not always available. 
In test-driven development (TDD) \cite{beck2002test}, test cases need to be defined before program implementation. 
In real-world software development, teams often rely solely on natural language requirements, as there are no pre-existing test cases or example programs. 
Furthermore, creating pre-defined test cases can be labor-intensive, requiring significant manual effort. 
In situations where LLMs must generate test cases independently, the effectiveness of existing methods remains uncertain.

Some approaches do not rely on pre-defined test cases. 
For instance, methods like CodeT \cite{chen2022codet} and MBR-Exec \cite{shi2022natural} generate both code and test cases using LLMs and filter the code based on the execution results of these test cases.
Similarly, CodeCoT \cite{huang2023codecot}, 
which utilizes chain-of-thought prompting \cite{brown2020language}, 
enables LLMs to generate code and test cases simultaneously and evaluate and refine the code based on these cases. 
Additionally, AgentCoder \cite{huang2023agentcoder} introduces a multi-agent system with a dedicated Tester agent to generate test cases for evaluating and repairing code.
However, studies such as \cite{chen2022codet, li2024large} highlight that LLM-generated test cases can contain significant errors, potentially leading to unreliable evaluations and incorrect feedback.
Moreover, many of these methods fail to verify the quality of the generated test cases. 
While approaches implement basic filtering mechanisms for test cases, other methods often use generated test cases directly for code evaluation without ensuring their correctness. 

There are also previous co-evolution methods based on code and test cases, which have primarily focused on code maintenance and program repair. 
For example, \cite{shimmi2022leveraging} recommends new test cases based on source code similarity, \cite{chi2024reaccept} identifies outdated test cases, and \cite{ruan2024evolutionary} constructs test oracles from bug reports using predefined templates. 
Similarly, \cite{arcuri2007coevolving} generates test cases based on predefined specifications to drive software evolution. 
However, these approaches depend heavily on pre-existing programs, test specifications, or templates, limiting their applicability in scenarios lacking such resources.

\begin{figure}[!t]
\centering
\includegraphics[width=3.2in]{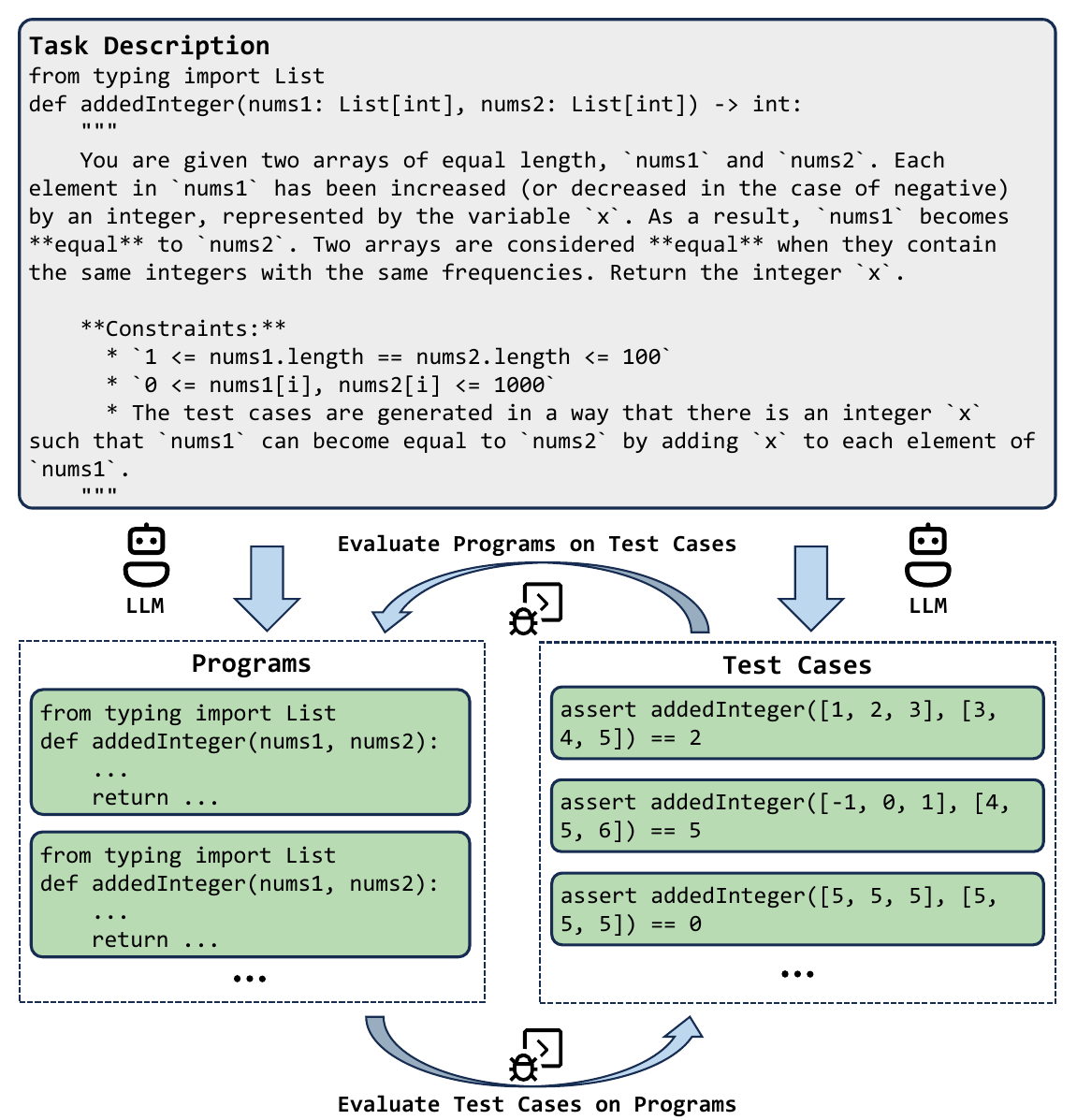}
\caption{
An overview of CoCoEvo. 
Function implementation is considered a program individual, and each assert statement is considered a test case individual. 
These two entities undergo an alternating co-evolutionary process through crossover evaluation. 
}
\label{fig:coevo_framework}
\end{figure}

To address these limitations, we propose \textbf{CoCoEvo}, an LLM-based co-evolution framework that enables the simultaneous evolution of programs and test cases to produce more accurate outcomes (illustrated in Fig. \ref{fig:coevo_framework}).
In CoCoEvo, both programs and test cases are generated by LLMs based on natural language problem descriptions.
No pre-defined programs, test cases, or templates are needed. 
The CoCoEvo framework consists of two alternating steps: program evolution and test case evolution. 
In program evolution, we introduce LLM-based crossover and mutation operators to generate program offspring. 
The current test case population is used to calculate program fitness. 
In test case evolution, we design an additional test case generation operator to produce test case offspring. 
Conversely, the current program population is used to calculate test case fitness. 
For programs, we implement a fitness calculation method inspired by CodeT \cite{chen2022codet}, which assesses agreement between programs and test cases.
For test cases, we adopt a multi-objective optimization approach to select those with high accuracy and strong discriminatory power, using the Pareto method \cite{deb2000fast}.
Additionally, we introduce a dynamic crossover rate scheduling mechanism, enabling broader exploration during early evolution stages and faster convergence later.

To evaluate the effectiveness of CoCoEvo, we collected a LeetCode-Contest dataset comprising relatively new programming problems that are unlikely to have been included in the training data of the evaluated LLMs.
We conducted experiments on four state-of-the-art models: GPT-4o-mini \cite{hurst2024gpt}, Qwen2.5-Coder-32B \cite{hui2024qwen2}, Llama-3.1-70B \cite{dubey2024llama}, and DeepSeek-V3 \cite{liu2024deepseek}. 
We compared with both basic code generation methods, repair-based methods, and agreement-based methods. 
The results demonstrate that CoCoEvo outperforms all existing methods.
Additionally, the experiments revealed that methods relying on pre-defined test cases for feedback experience significant performance degradation when adapted to self-generated test cases. 
This highlights the limitations of such methods in scenarios lacking pre-defined test cases. 
Furthermore, we conducted in-depth comparative experiments and discussions, which validated the effectiveness of the optimization strategies employed in CoCoEvo. 

In summary, our main contributions are as follows:
\begin{enumerate}
\item{
Co-evolution framework.
We propose a novel LLM-based co-evolution framework for programs and test cases, 
including specialized LLM-based operators for offspring generation.
}
\item{
Optimized Evolution Techniques.
For program evolution, 
we design a dynamic crossover rate scheduler that enhances exploration early on and accelerates convergence later. 
For test cases, we develop a multi-objective optimization method for fitness calculation and new population selection.
}
\item{
Experimental Validation.
We introduce the LeetCode-Contest dataset and conduct experiments using four state-of-the-art LLMs: GPT-4o-mini, Qwen2.5-Coder-32b, Llama-3.1-70B, and DeepSeek-V3.
The experimental results demonstrate that CoCoEvo achieves state-of-the-art performance, surpassing existing methods.
Additionally, we conduct comparative experiments to evaluate the utility of each module within the framework, further highlighting the effectiveness of our approach.
}
\end{enumerate}
The remainder of this article is organized as follows. 
Section \ref{sec:related-work} discusses the related work.
In Section \ref{sec:method}, the design details and implementation methods of the CoCoEvo are described. 
Sections \ref{sec:experiments} and \ref{sec:results} present the experimental settings and discussions, respectively.
Finally, Section \ref{sec:conclusion} concludes this paper and discusses the limitations and future work.

\section{Related Work}
\label{sec:related-work}
\subsection{Enhancing Code Generation Accuracy with Test Case}

Integrating test cases is now crucial for improving LLM-generated code. Key methodologies use them to filter, evaluate, and refine programs, enhancing performance and reliability.

Li et al. \cite{li2022competition}'s Sampling+Filtering approach samples numerous candidate programs and scores them against a given test set. 
Chen et al. \cite{chen2022codet} use LLMs to co-generate code and tests, selecting superior programs via a dual agreement metric. 
Shi et al. \cite{shi2022natural} use test execution results to assess program similarity, guiding program selection.

Beyond evaluation, test-case results drive iterative repair and refinement of generated code.
Reflexion \cite{shinn2024reflexion} converts test failures into verbal feedback for program repair. 
Similarly, Olausson et al. \cite{olausson2023self} evaluate LLMs' capabilities to repair programs using failed test feedback. 
Zhang et al. \cite{zhang2023self} allow LLMs to analyze test execution results to revise erroneous code.
Zhong et al. \cite{Zhong0S24ldb} extend this idea by enabling step-by-step debugging.
CodeCoT \cite{huang2023codecot} uses chain-of-thought prompting \cite{brown2020language} to sequentially generate code and tests, iteratively refining the code with test feedback. 
Zhang et al. \cite{zhang2024pair} introduce a pair programming framework where test feedback informs repair strategies and planning.

The generation of test cases differs across these methodologies.
While CodeT \cite{chen2022codet} and CodeCoT \cite{huang2023codecot} use LLMs exclusively for test generation, Reflexion \cite{shinn2024reflexion} employs them mainly for simple programming tasks. 
However, most methods rely on predefined test cases to filter, refine, and select programs.

\subsection{Automated Test Case Generation}
Early works, such as \cite{fraser2011evosuite}, leveraged search-based methods to generate test cases. 
More recently, research has shifted toward using LLMs for this task. 
Studies such as \cite{siddiq2024using}, \cite{guilherme2023initial}, and \cite{schafer2023empirical} have explored the effectiveness of LLMs in automating program testing. 
Similarly, \cite{alshahwan2024automated} investigates methods for enhancing existing test cases using LLMs. 
Others, like TOGA \cite{dinella2022toga}, TOGLL \cite{hossain2024togll}, and \cite{hossain2023neural},  generate test oracles from predefined templates. 
Meanwhile, \cite{takerngsaksiri2024tdd} trained a deep reinforcement learning model for text-to-test case generation. 
The application of LLMs in software development extends to generating test cases for various scenarios. 
For instance, \cite{zhang2023well} employs GPT-4 to generate security tests for identifying vulnerabilities, while \cite{schafer2023empirical} uses LLMs to assist in generating unit tests for automatic software testing. 
Similarly, \cite{kang2023large} demonstrates the use of LLMs to create test cases aimed at reproducing general software bugs. 
LLMs are also employed to improve the code accuracy through test case generation. 
CodeT \cite{chen2022codet}, for example, utilizes LLMs with zero-shot prompts to directly generate test cases. 
Reflexion \cite{shinn2024reflexion} introduces a Test Agent powered by LLMs for test case generation.

\subsection{Evolution with LLMs for Code Generation}
The integration of LLMs with evolutionary algorithms has been extensively explored to enhance code generation. 
Bradley et al. \cite{bradley2024openelm} introduced OpenELM, an open-source Python library that designs specialized evolutionary operators for code generation. 
Chen et al. \cite{chen2023evoprompting} employed LLMs as adaptive operators for evolutionary neural architecture search to optimize network designs.
Liu et al. \cite{liu2023algorithm} proposed using LLMs to evolve optimization algorithms, automating the processes of initialization, selection, and mutation. 
Liu et al. \cite{liu2024evolution} introduced a framework integrating LLMs with evolutionary computation to co-evolve heuristic descriptions and executable code.
Ma et al. \cite{ma2023eureka} combined LLMs with evolutionary search to generate and refine RL reward functions via a reward-reflection mechanism.
Meyerson et al. 
\cite{meyerson2024language} used few-shot prompting to create a domain-agnostic crossover operator by concatenating parent solutions into one prompt. 
Similarly, Romera et al. \cite{romera2024mathematical} integrated LLMs with evolutionary algorithms to create context-aware program mutations. 
Ye et al. \cite{ye2024reevo}'s ReEvo combines evolutionary search and LLM-driven reflections to generate heuristics via verbal optimization gradients. 
AutoTest \cite{duan2024autotest} embeds LLM co-generation of code and test cases in an evolutionary loop for improved optimization.

Additionally, various LLM-based co-evolution methods have been proposed. 
For instance, Chi et al. \cite{chi2024reaccept} identifies outdated test cases by co-evolving production and test code. 
Ruan et al. \cite{ruan2024evolutionary} constructs test oracles from bug reports using predefined templates. 
Chi et al. \cite{chi2024reaccept} used LLMs with dynamic validation to automate production and test code co-evolution, identifying obsolete tests to enhance software quality and lower maintenance overhead. 
TestART \cite{gu2024testart} enhances LLM unit testing by automatically generating and repairing tests, using template-based fixes and coverage feedback to improve accuracy.

\section{Methods}
\label{sec:method}
\subsection{Framework Overview}

\begin{algorithm}[H]
\caption{Pseudocode of CoCoEvo}
\label{alg:cocoevo}
\begin{algorithmic}[1]
\REQUIRE An $LLM$, program population size $N^P$, max iteration rounds $max\_iter$
\ENSURE Program $p_{best}$ with the highest fitness
\STATE // Initialization
\STATE Initialize program population $P$ and test case population $T$ with the $LLM$;
\STATE Cross evaluate $P$ and $T$ to obtain $F^P$;
\STATE Obtain the program with the highest fitness as $p_{best}$;
\STATE // Co-evolution loop
\FOR{$r = 2$ \TO $max\_iter$}
    \STATE // Program evolution
    \STATE Calculate $x_r$ via the cosine scheduler;
    \STATE Crossover operations $Nc^P_r \gets \lfloor x_r \times N^P \rfloor$;
    \STATE Mutation operations $Nm^P_r \gets N^P - Nc^P_r$;

    \STATE Program offspring $P' \gets \emptyset$;
    \FOR{$i = 1$ \TO $Nc^P_r$}
        \STATE Select programs $p_1$, $p_2$ from $P$ with the binary tournament algorithm;
        \STATE Perform crossover on $p_1$ and $p_2$ with the $LLM$ to obtain $p'$;
        \STATE $P' \gets P' \cup \{p'\}$;
    \ENDFOR
    \FOR{$i = 1$ \TO $Nm^P_r$}
        \STATE Randomly select program $p$ from $P$;
        \STATE Perform mutation on $p$ with the $LLM$ to obtain $p'$;
        \STATE $P' \gets P' \cup \{p'\}$;
    \ENDFOR

    \STATE Evaluate $P'$ on $T$ , calculate $F^P$;
    \STATE $P \gets P \cup P'$;
    \STATE Select $N^P$ programs from $P$ with the highest fitness;
    \STATE Update the program $p_{best}$ with the highest fitness;

    \STATE // Test case evolution
    \STATE Generate line coverage information $feedback$ based on $T$ and $p_{best}$;
    \STATE Generate additional test cases $T'$ based on $T$ and $feedback$ with the $LLM$;
    \STATE $T \gets T \cup T'$;
    \STATE Cross evaluate $P$ and $T$ to obtain $F^P$, $Conf^{T}$, and $Disc^{T}$;
    \STATE Select the new test case population $T$ based on $Conf^T$ and $Disc^T$ with the Pareto method;
    \STATE Update the program $p_{best}$ with the highest fitness;
\ENDFOR
\RETURN $p_{best}$.
\end{algorithmic}
\end{algorithm}

\begin{figure*}[!htpb]
\centering
\includegraphics[width=6.6in]{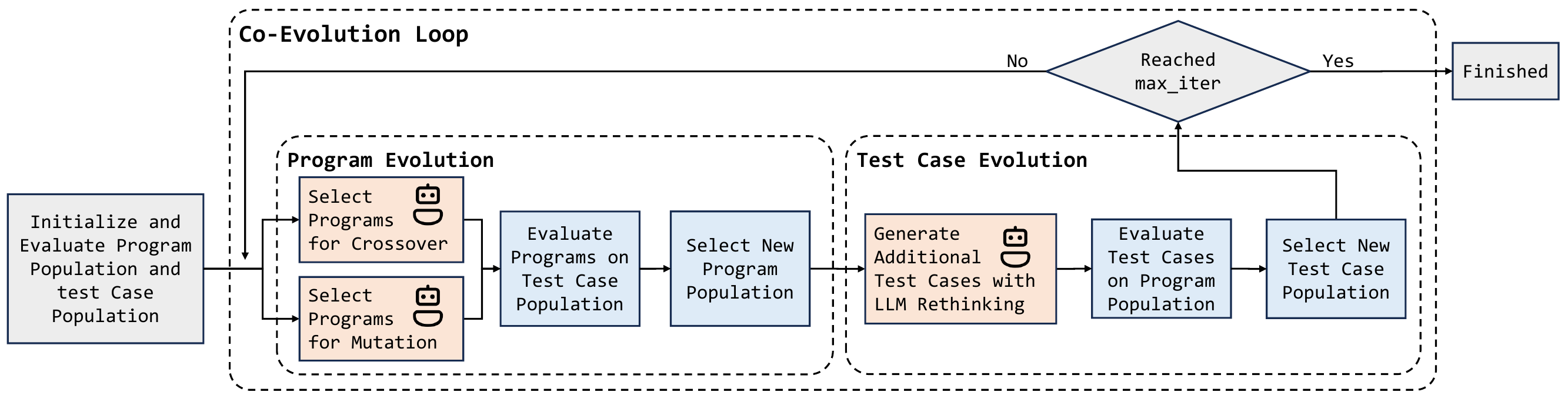}
\caption{
An overview of the co-evolution loop.
The workflow consists of alternating phases of program evolution and test case evolution.
In the program evolution phase, the test case population is used to evaluate the fitness of programs, and the offspring programs are generated through crossover and mutation operators.
In the test case evolution phase, the program population is used to evaluate the fitness of test cases, and the offspring test cases are produced by generating additional test cases.
Rectangles with an orange background indicate the steps involving the use of LLMs.}
\label{fig:coevo_loop}
\end{figure*}

An overview of CoCoEvo is presented in Fig. \ref{fig:coevo_loop}. 
The process begins by randomly initializing both the program and test-case populations, using LLMs to generate the initial programs and test cases.
The co-evolution process follows a loop consisting of two alternating steps: program evolution and test case evolution, each of which includes a \emph{Generation-Evaluation-Selection} sequence.
During program evolution, the test case population is used to evaluate programs and calculate their fitness. 
Conversely, during test case evolution, the program population is utilized to evaluate test cases.

For programs, we introduce LLM-based crossover and mutation operators for offspring generation.
For test cases, we implement an LLM-based additional test case generation operator.
Once the preset number of iterations is reached, the co-evolution loop terminates, and the program with the highest fitness score is returned as the final answer.
Algorithm \ref{alg:cocoevo} provides the pseudo-code detailing the CoCoEvo process.
Further implementation details for each algorithm step are elaborated in subsequent sections.

\subsection{Program Evolution}

\begin{figure}[htpb]
\centering
\includegraphics[width=1.0in]{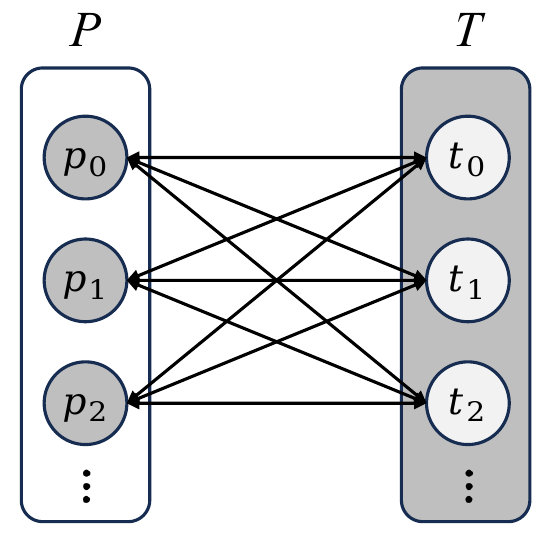}
\caption{
Illustration of the cross-evaluation process across programs and test cases.
Each program and each test case are executed independently, producing either a pass or fail outcome.
In the figure, $P$ denotes programs, and $T$ denotes test cases.
Moreover, $p_0$, $p_1$, $p_2$, ... represent each program, while $t_0$, $t_1$, $t_2$, ... represent each test case.
}
\label{fig:cross}
\end{figure}

\textbf{Program evaluation.}
During evaluation, programs and test cases are cross-evaluated, as illustrated in Fig. \ref{fig:cross}. 
The outcomes are represented by a matrix $M$:
\begin{equation}
\label{eq:cross}
M_{i,j} = \begin{cases} 
1, & \text{\(p_i\) passed \(t_j\)} \\
0, & \text{\(p_i\) failed \(t_j\)}
\end{cases}
\end{equation}
where $p_i$ and $t_j$ represent the $i$‑th program and the $j$‑th test case, respectively.

We design a Confidence metric to evaluate program fitness, drawing inspiration from CodeT \cite{chen2022codet}. 
Programs that pass the same set of test cases are grouped into a set $Z^P$, while the corresponding test cases are grouped into a set $Z^T$.
The Confidence of program $i$ is defined as
\begin{equation}
\label{eq:code-codet}
Conf^P(p_i) = \sqrt{|Z^P|} \times |Z^T|
\end{equation}
where the size of $Z^P$ and $Z^T$ are denoted by $|Z^P|$ and $|Z^T|$, respectively.
This Confidence score evaluates the likelihood of a program being correct.
The fitness of program $i$ is defined as
\begin{equation}
\label{eq:code-fitness}
F^P(p_i) = Conf^P(p_i)
\end{equation}
where a higher fitness score indicates a higher probability of correctness for the program.

\textbf{Crossover Rate Scheduler.}
When leveraging LLMs to perform program crossover and mutation, the crossover operation typically synthesizes the strengths of disparate programs but may drive the population toward premature convergence, whereas the mutation operation often introduces novel features yet cannot guarantee evolution in a more desirable direction.
Therefore, we propose a dynamic crossover rate scheduler that adaptively adjusts the number of crossover and mutation operations.
We define the crossover rate as the proportion of offspring programs generated through the crossover operator. 
In the early stages, a lower crossover rate is applied to promote exploration, 
and the crossover rate increases in later stages to refine convergence. 
This enables extensive exploration of the search space during the early stages of evolution, while facilitating rapid convergence toward superior solutions in later generations.

Unlike traditional random mutation methods, the LLM-based program mutation operator leverages LLMs for code generation, which could retain key features of high-quality programs. 
This allows for a higher mutation rate without significant loss of desirable traits. 
Thus, we adopt the cosine annealing method \cite{LoshchilovH17cosine} for scheduling crossover rate:
\begin{equation}
\label{eq:cosine}
x_r = x_{final} + \frac{1}{2}( x_{init} - x_{final})(1+cos(\frac{\pi (r-1)}{max\_iter - 1}))
\end{equation}
where $x_r$ denotes the crossover rate at the $r$-th iteration, $x_{init}$ and $x_{final}$ represent the initial crossover rate and the final crossover rate, which are set to $0.0$ and $1.0$, respectively.
During the iteration process, $r$ starts at $1$ and ends at $max\_iter$. 
In the formula, both $r$ and $max\_iter$ are decremented by $1$, since in the first iteration ($r = 1$) the program population is doing initialization and no crossover or mutation operations are performed.

This formulation ensures that the crossover rate increases smoothly from $0$ to $1$.
In each generation, $N^P$ offspring programs are produced.
The number of crossover operations is determined by multiplying the crossover rate by $N^P$ and taking the floor of the result.
The number of mutation operations is equal to $N^P$ minus the number of crossover operations, as calculated by the following equations:
\begin{equation}
\label{eq:rate}
\begin{cases} 
Nc^P_r = \lfloor x_r N^P \rfloor \\
Nm^P_r = N^P - Nc^P_r
\end{cases}
\end{equation}
where $N^P$ is the program population size,
$Nc^P_r$ and $Nm^P_r$ represent the number of program crossover and mutation operations in the $r$-th iteration, respectively.

\begin{figure}[!t]
\centering
\includegraphics[width=2.0in]{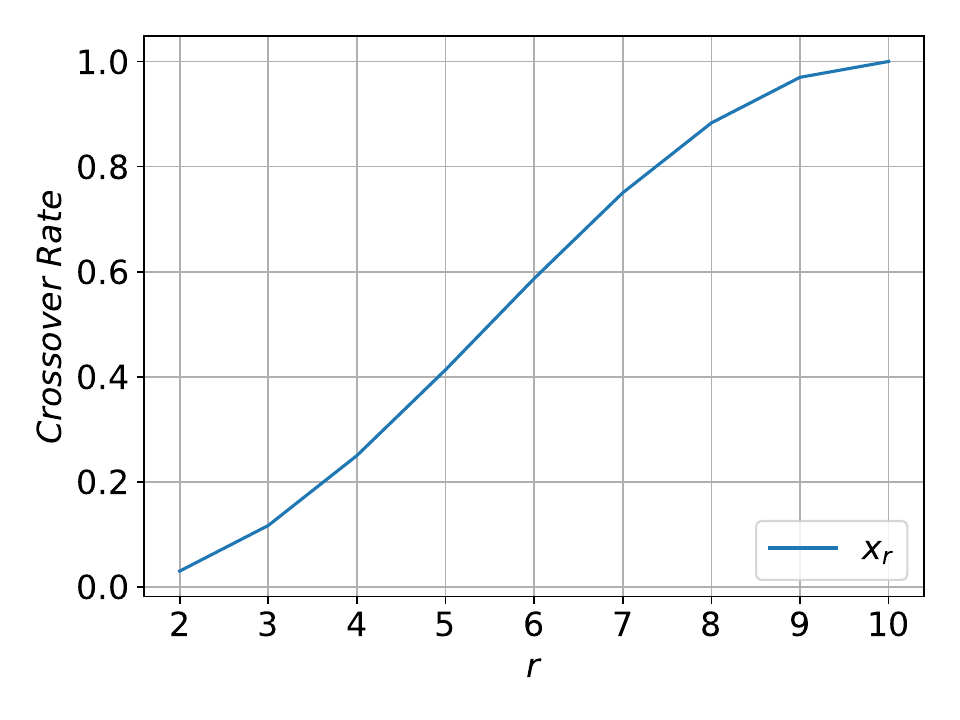}
\caption{
Illustration of the variation of the crossover rate scheduled based on the cosine annealing method.
The crossover rate denotes the proportion of offspring generated through crossover operations during the reproduction process. 
$r$ denote the $r$-th iteration, and $x_r$ represent the crossover rate at that iteration.
In the figure, $r$ starts from $2$ because when $r=1$ only population initialization takes place and no crossover operations are required.
}
\label{fig:cosine}
\end{figure}

Fig. \ref{fig:cosine} presents an example of the variation in crossover rate over $10$ iterations.
Since when $r=1$, the population is randomly initialized, the figure illustrates changes in the crossover rate from $r=2$ to $10$, with the parameter $max\_iter$ set to $10$.
This approach ensures a smooth transition between exploration (early stages) and convergence (later stages), 
enabling effective optimization of the solution space.

\subsection{Test Case Evolution}
\label{sec:test-evo}

\textbf{Test case evaluation.}
The evaluation of test cases involves two key metrics: \emph{Confidence} and \emph{Discrimination}.
Confidence evaluates the probability that a test case is correct, and discriminative power evaluates a test case’s ability to distinguish between programs. 
The Confidence of test case $j$ is calculated as
\begin{equation}
\label{eq:test-conf}
Conf^T(t_j) = \frac {1} {N^P} \sum_{i=1}^{N^P} {M_{i,j} F^P(p_i)}
\end{equation}
where $N^P$ represents the size of the program population, $F^P(p_i)$ represents the fitness of program $i$, and $M_{i,j}$ is the evaluation matrix indicating whether $p_i$ passed $t_j$.
Intuitively, Confidence measures the degree of agreement between the program population and the test cases, with higher program fitness contributing more weight to the degree of agreement.

Simple test cases may fail to differentiate between correct and erroneous programs, as both types may pass them. 
To address this, we introduce a Discrimination metric. 
First, the pass rate of test case $j$ across the program population is defined as
\begin{equation}
\label{eq:test-p}
pr^T_j = \frac {1} {N^P} \sum_{j=1}^{N^P} {M_{i,j}}
\end{equation}
then the Discrimination of test case $j$ is represented by the entropy of the pass rate:
\begin{equation}
\label{eq:test-disc}
Disc^T(t_j) = -pr^T_j log_{2} pr^T_j - (1-pr^T_j) log_{2} (1-pr^T_j).
\end{equation}

Test cases achieve higher Discrimination when their pass rate approaches $0.5$, as they better differentiate among the program population. 
This indicator assesses whether the test cases can exert evolutionary pressure on the program population.

\begin{figure}[!ht]
\centering
\includegraphics[width=1.8in]{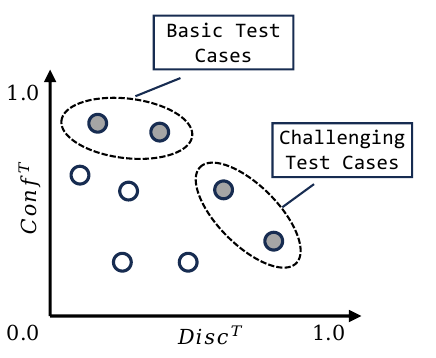}
\caption{
Illustration of the Pareto front constructed based on the $Conf^T$ and $Disc^T$.
The $Conf^T$ and $Disc^T$ represent the Confidence of the test cases and the Discrimination of the test cases.
The Pareto front is computed from the two indicators, and these test cases are selected as the next-generation population.
}
\label{fig:pareto}
\end{figure}

\textbf{Test case selection.}
The selection of test cases is performed using the Pareto approach for multi-objective optimization \cite{deb2000fast}.
The Pareto frontier, also known as the Pareto front or Pareto boundary, represents the set of non-dominated solutions in a multi-objective optimization problem.
A solution is considered Pareto optimal if no other feasible solution improves one objective without causing a deterioration in at least one other.
Formally, given a vector-valued objective function $f(x)=[f_1(x), f_2(x),...,f_n(x)]$, a solution $x$ is Pareto optimal if there does not exist another solution $y$ such that $f_i(x) \leq f_i(y)$ for all $i=1,...,n$, with at least one strict inequality.

In test case selection, the Pareto front is constructed based on the two metrics, $Conf^T$ and $Disc^T$, which represent the Confidence of test cases and the Discrimination of test cases.
The Pareto front is illustrated in Fig. \ref{fig:pareto}.
Test cases with high $Conf^T$ and low $Disc^T$ represent basic test cases that cover common scenarios, with a high probability of being correct.
Test cases with high $Disc^T$ represent more challenging cases that are effective at exposing errors.
During the selection process, preference is given to individuals that perform well on both indicators. 
This strategy aims to: (1) preserve the correctness of the test case population, ensuring accurate evaluation of the program, and (2) retain test cases with high discriminative power to maintain selection pressure on the program population. 
We also noticed that there are some test cases with very low $Conf^T$ that sometimes have a relatively high $Disc^T$.
To avoid adverse effects of these test cases on program evolution, the test cases with $Conf^T$ lower than the average in the selection results will be filtered out.

\subsection{LLM-based Evolution Operators}

\begin{figure}[!t]
\centering
\includegraphics[width=3.4in]{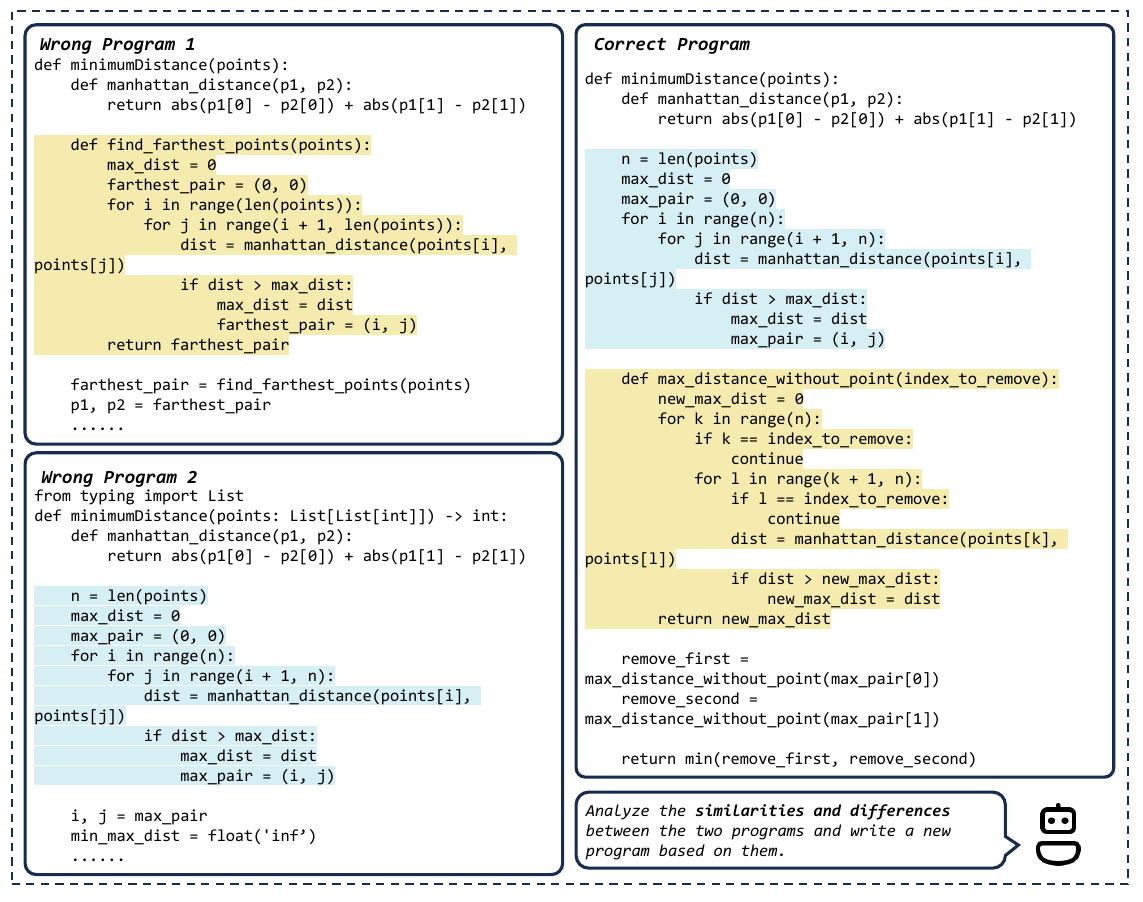}
\caption{
Illustration of LLM-based program crossover.
The code blocks with the same background color indicate the segments of the parent program that are referenced by the LLM during crossover.
}
\label{fig:crossover}
\end{figure}

\textbf{Program crossover operator.} 
We utilize LLMs to perform program crossover,
as illustrated in Fig. \ref{fig:crossover}. 
The process involves the following steps:
\begin{enumerate}
\item{
Selection. Two programs are selected as crossover parents using the binary tournament selection method based on their fitness scores.
}
\item{
Crossover. The selected programs are combined into a prompt. The LLM is instructed to analyze their similarities and differences and generate a new program that merges useful elements from both.
}
\end{enumerate}

For example, in Fig. \ref{fig:crossover},
two incorrect programs are sent to the LLM.
By analyzing their structures, the LLM combines useful fragments from each and generates a new program that resolves the errors.

\textbf{Program mutation operator.}
The LLM-based mutation operator intelligently rewrites a program to achieve the same functionality using a different approach.
The steps are as follows:
\begin{enumerate}
\item{
Selection. A program is randomly selected from the current population as the mutation parent.
}
\item{
Mutation. The parent program is included in a prompt,
and the LLM is instructed to rewrite it using an alternative implementation method.
}
\end{enumerate}

\begin{figure}[!t]
\centering
\includegraphics[width=3.4in]{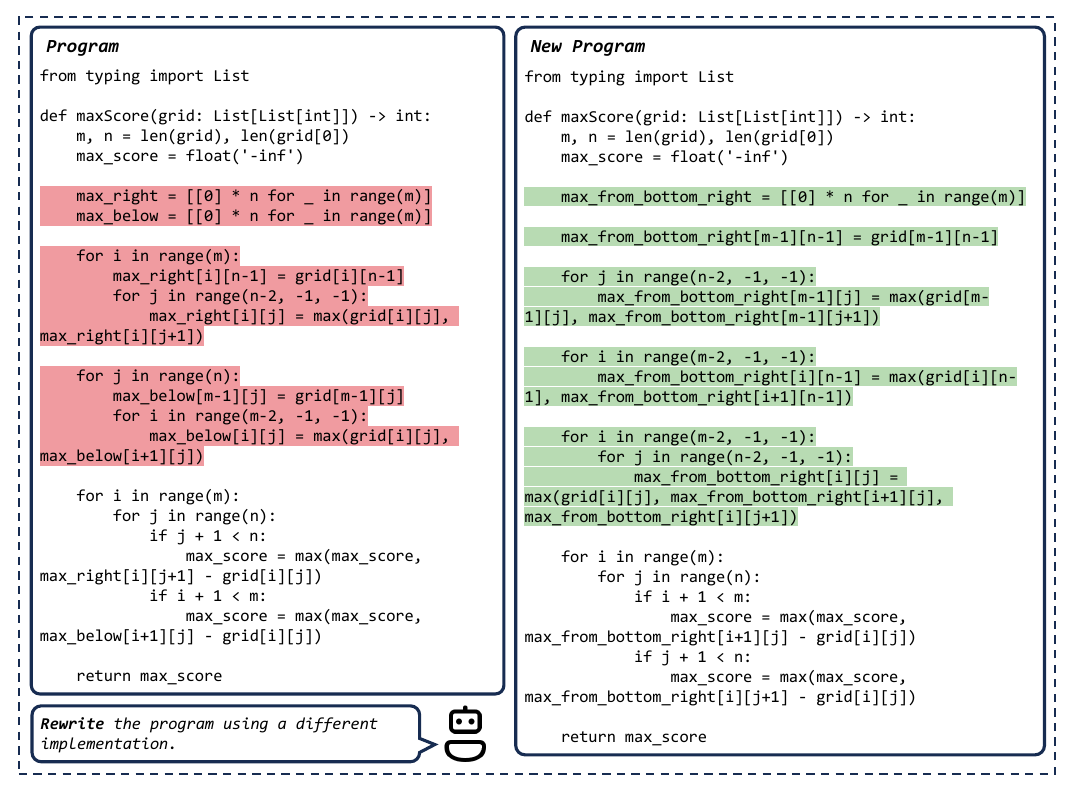}
\caption{
Illustration of LLM-based program mutation.
During mutation, the LLM rewrote certain program segments,
replacing the red background part of the program on the left with the green background part on the right,
thereby correcting errors in the parent program.
}
\label{fig:mutation}
\end{figure}

For example, Fig. \ref{fig:mutation} demonstrates a mutation scenario where the task is to find the maximum difference between adjacent elements in a 2D grid.
The parent program has a flaw—it only considers elements in the same row or column.
The LLM analyzes the problem and rewrites the program to include diagonal relationships, 
successfully resolving the flaw.

\begin{figure}[!t]
\centering
\includegraphics[width=3.0in]{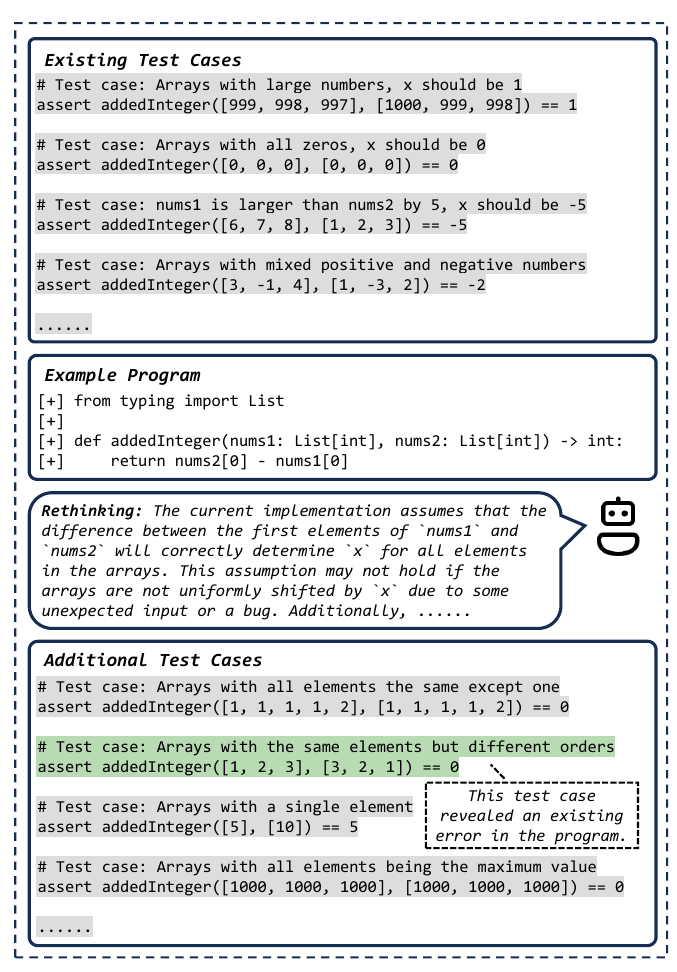}
\caption{
Illustration of LLM-based rethinking and additional test case generation.
The LLM identified scenarios that were not covered by the current test case population and generated an edge test case (the test case with the green background).
}
\label{fig:test_gen}
\end{figure}

\textbf{Additional test cases generation operator.}
LLMs are also employed to generate additional test cases,
as shown in Fig. \ref{fig:test_gen}. 
The process includes the following steps:
\begin{enumerate}
\item {
Information Construct.
The prompt contains the following information: The entire existing population of test cases, the best program in the program population, and the line coverage information on it, where covered lines are marked as ``[+]'' and uncovered lines as ``[-]''.
}
\item {
Rethinking and Generation.
Based on the information, the LLMs are required to generate additional test cases:
If there are uncovered lines, the LLMs generate test cases to achieve full line coverage.
If all lines are already covered, the LLMs analyze the existing test cases to identify untested boundary conditions that could reveal potential program flaws.
}
\end{enumerate}

For instance, in Fig. \ref{fig:test_gen}, the LLM identifies an untested boundary scenario: the input list $nums2$ is not in ascending order.
It then generates a new test case with a non-ascending $nums2$, exposing the flaw in the provided program.

\section{Experiments}
\label{sec:experiments}

\subsection{Dataset}
We collected a dataset of $80$ programming problems from the weekly and biweekly contests hosted on the LeetCode platform \footnote{\url{https://leetcode.com/}}, forming the LeetCode-Contest dataset.
In the dataset, each problem includes a prompt comprising the function header and a natural language description in the form of a docstring, along with program solutions and approximately $644$ ground truth test cases per problem. 
These comprehensive ground truth test cases can closely simulate the real-world scenario of submissions on LeetCode, making our evaluation more convincing.
To ensure data integrity and avoid potential leakage, all selected problems were released after March 2024.
To more accurately assess the ability of LLMs to solve programming challenges, 
we took measures to prevent plagiarism of example test inputs and outputs in test case generation.
Specifically, we removed all test input-output samples from the problem descriptions, 
retaining only the natural language problem statements, the function header, and the associated constraints.
The detailed information of the LeetCode-Contest dataset is summarized in Table \ref{tab:leetcode}.

\begin{table}[!t]
\caption{
Illustration of the detailed information of the LeetCode-Contest dataset.
\label{tab:leetcode}
}
\centering
\begin{tabular}{lc}
\toprule
Problems            & 80 \\
\midrule
Average Test Cases  & 644.40 \\
\midrule
Selected Contests   & 
\makecell{
weekly-contest-402, weekly-contest-401, \\
biweekly-contest-132, weekly-contest-400, \\
weekly-contest-399, biweekly-contest-131, \\
weekly-contest-398, weekly-contest-397, \\
biweekly-contest-130, weekly-contest-396, \\
weekly-contest-395, biweekly-contest-129, \\
weekly-contest-394, weekly-contest-393, \\
biweekly-contest-128, weekly-contest-392, \\
weekly-contest-391, biweekly-contest-127, \\
weekly-contest-390, weekly-contest-389
} \\
\bottomrule
\end{tabular}
\end{table}

\subsection{Models}
We utilized four widely adopted and powerful LLMs for our experiments: GPT-4o-mini \cite{hurst2024gpt}, Qwen2.5-Coder-32B \cite{hui2024qwen2}, Llama-3.1-70B \cite{dubey2024llama}, and DeepSeek-V3 \cite{liu2024deepseek}.
For GPT-4o-mini, we accessed the API via the OpenAI platform \footnote{\url{https://platform.openai.com/}}.
For Qwen2.5-Coder-32B and Llama-3.1-70B, we used the APIs available on the DeepInfra platform \footnote{\url{https://deepinfra.com/}}.
For DeepSeek-V3, we utilized the API provided by the DeepSeek platform \footnote{\url{https://platform.deepseek.com/}}.

\subsection{Compared Baselines}
We selected several of the most prominent and effective methods in the field of code generation with LLMs as our comparison baselines:
\begin{enumerate}
\item{
Sampling. 
A basic method that generates a large number of code samples and randomly selects one as the final answer. 
}
\item{
Sampling+Filtering \cite{li2022competition}. 
An enhancement of the Sampling method, where the generated codes are evaluated using test cases. 
The code with the highest score is selected as the final answer. 
}
\item{
Self-Repair \cite{olausson2023self}.
This method generates code randomly, and then feeds execution results on test cases back to the LLM, which repairs the code accordingly. 
}
\item{
Reflexion \cite{shinn2024reflexion}.
Error information from test case executions is used to guide the LLM in providing reflection and repair suggestions. The code is then repaired based on these suggestions.
}
\item{
INTERVENOR \cite{wang2023intervenor}.
Utilizes a Teacher-Student multi-agent framework. The Teacher agent analyzes error information from test case execution and provides repair suggestions, while the Student agent applies the fixes.
}
\item{
CodeCoT \cite{huang2023codecot}.
Based on the chain-of-thought paradigm \cite{brown2020language}, this method requires the LLM to generate both code and test cases simultaneously. Feedback from execution errors is then used to repair the code.
}
\item{
AgentCoder \cite{huang2023agentcoder}.
Implements two LLM-based agents: a Programming agent for code generation and repair, and a Testing agent for test case generation.
}
\item{
MBR-Exec \cite{shi2022natural}.
Generates both code and test cases using LLMs.
A loss function evaluates execution similarity based on the outputs of the generated code when run on test inputs.
Codes with higher execution divergence receive higher losses, and the code with the lowest loss is selected as the final answer.
}
\item{
CodeT \cite{chen2022codet}.
Generates code and test cases with LLMs, 
then performs clustering based on how the codes perform on the test cases.
The clustering score is calculated as the product of the number of codes in a cluster and the number of test cases passed.
}
\end{enumerate}

For the baselines (Sampling+Filtering, Self-Repair, Reflexion, INTERVENOR) that originally relied on pre-defined test cases for program evaluation, we adapt them to our setting by replacing the pre-defined test cases with LLM self-generated test cases.

\subsection{Performance Indicators}
The ground truth test cases in the dataset are used to evaluate the accuracy of the generated program. 
A code is considered correct if it passes all the ground truth test cases. 
The performance of the approaches on the dataset is measured using the pass@1 metric \cite{kulal2019spoc}.
To calculate the pass@1 metric, each method submits a single program as the final answer for each problem.
The pass@1 metric represents the percentage of problems successfully solved. 
Additionally, for further analysis of the CoCoEvo approach, 
the ground truth program solutions are used to evaluate the correctness of the generated test cases.
A generated test case is considered correct if it executes successfully on the ground truth solution.

\subsection{Experimental Settings}
\label{subsec:experimental-settings}

Following the comparison settings used in prior studies \cite{olausson2023self, shi2022natural, chen2022codet}, we compare the performance of different methods by constraining the number of code generation attempts made by the LLMs to be equal. 
Based on the common limitations of these approaches and taking into account the cost of invoking LLMs, we limit the number of LLM code generation calls to $100$. 
For methods that require test case generation, we impose a similar constraint by limiting the number of test case generation calls made to the LLMs to $10$, with a maximum of $10$ test cases extracted per generation. 
For CoCoEvo, we adopt a balanced configuration by setting both the population size $N^P$ and the number of iterations $max\_iter$ to $10$.
Comparative experiments with alternative population sizes and detailed parameter settings of baselines and CoCoEvo can be found in the supplementary material. 
Since different code generation strategies may require varying numbers of tokens, we provide a detailed analysis of token consumption in Section \ref{subsec:analysis-of-token-usage}. 
We observed that when calculating the pass@1 metric, 
if multiple codes achieve the same score, one code will be randomly selected as the final result. 
This introduces a certain degree of randomness to the evaluation process. 
To mitigate this, for all methods, we independently and randomly repeated the ``selection $\rightarrow$ evaluation'' process $5$ times and reported the averaged results.

\section{Results and Discussion}
\label{sec:results}
\subsection{Comparison of Accuracy}

\begin{table*}[!ht]
\caption{
Illustration of pass@1 performance across all methods on the LeetCode-Contest dataset.
\label{tab:pass1}
}
\centering
\begin{threeparttable}
\begin{tabular}{lcccccc}
\toprule
Method & GPT-4o-mini & Qwen2.5-Coder-32B & Llama-3.1-70B & DeepSeek-V3 \\
\midrule

Sampling
& 35.25 & 44.00 & 32.00 & 69.00  \\
Sampling+Filtering\tnote{*}
& 37.50 & 44.50 & 31.00 & 68.25  \\
\midrule

Self-Repair\tnote{*}
& 33.00 & 42.00 & 29.25 & 60.75  \\
Reflexion\tnote{*}
& 37.50 & 48.75 & 25.00 & 62.50  \\
INTERVENOR\tnote{*}
& 27.75 & 20.25 & 20.25 & 47.50  \\
\midrule

CodeCoT
& 31.25 & 36.50 & 19.50 & 63.50 \\
AgentCoder
& 31.25 & 41.75 & 22.00 & 60.25  \\
\midrule
MBR-Exec
& 33.75 & 45.00 & 38.75 & 70.00 \\
CodeT
& 46.25 & 47.50 & 41.25 & 72.50  \\
\midrule

\textbf{CoCoEvo}
& \textbf{49.75} & \textbf{55.75} & \textbf{45.00} & \textbf{76.25} \\
\bottomrule
\end{tabular}
\begin{tablenotes}
\item[*] Denotes that the methods were originally designed to operate with pre-defined test cases.
For this work, we adapted these methods to utilize test cases generated by the LLMs.
\end{tablenotes}
\end{threeparttable}

\end{table*}

In this section, we present the results of the accuracy comparison for the methods.
Table \ref{tab:pass1} reports the pass@1 metric for each method across different LLMs on the LeetCode-Contest dataset.
As shown in Table \ref{tab:pass1}, our proposed CoCoEvo method achieves the highest performance across all LLMs, demonstrating its superior ability to generate accurate code. 
This success can be attributed to the co-evolution framework employed by CoCoEvo, which facilitates a more effective iterative refinement between programs and test cases. 

An interesting observation concerns the code repair based methods such as Self-Repair, Reflexion, and INTERVENOR. 
When these methods utilize test cases generated by LLMs instead of pre-defined ones, their effectiveness diminishes significantly. 
In some instances, their performance is even worse than that of the basic Sampling approach. 
This is primarily due to the faulty LLM-generated tests mislead the repair process, leading to suboptimal code modifications and degraded overall performance.
Similarly, Sampling+Filtering yields only a slight improvement over plain Sampling under these conditions, highlighting these approaches’ dependence on reliable test suites.

Furthermore, methods that leverage the agreement between programs and test cases, such as MBR-Exec and CodeT, outperform both Sampling and repair-based methods.
Among these, CodeT stands out, delivering consistent improvements in pass@1 across all LLMs, suggesting that mutual validation of code and test cases is a promising strategy for improving generation accuracy.

\begin{figure*}[!ht]
\centering
\includegraphics[width=6.6in]{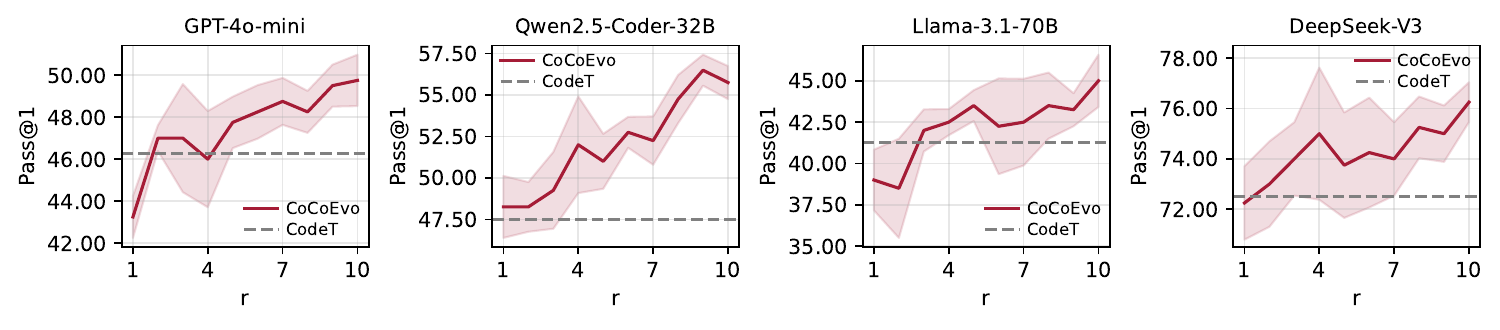}
\caption{
Illustrates the pass@1 performance during the evolutionary process.
The shaded region in the graph indicates the data distribution, expressed as the standard deviation. 
The gray dashed line represents the final performance achieved by the CodeT method.
In the figures, $r$ represents the iteration round. 
}
\label{fig:gens}
\end{figure*}

Fig. \ref{fig:gens} illustrates the changes of pass@1 during the evolution process.
While fluctuations in performance are observed,
the method demonstrates consistently strong performance on average. 
The figure shows that although the CoCoEvo method starts with a relatively low pass@1,
its performance steadily improves over the course of evolution iterations, eventually surpassing that of the CodeT method. 
This demonstrates the effectiveness of the CoCoEvo method. 
By alternating the evolution of programs and test cases, the program population progressively converges toward the correct solution.

\begin{figure*}[!ht]
\centering
\includegraphics[width=6.6in]{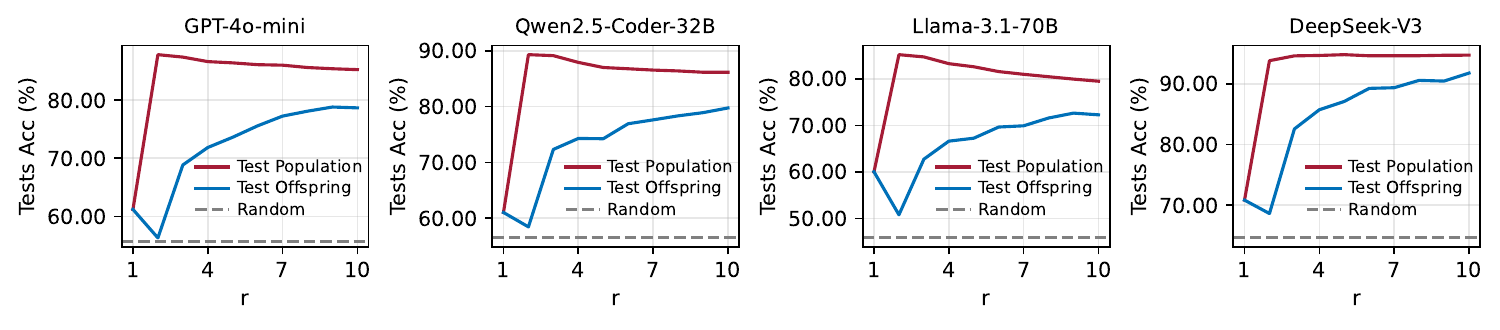}
\caption{
Illustrates the accuracy of the test case population and offspring during the evolutionary process, with the accuracy of the randomly generated test cases as a baseline.
The variable $r$ refers to the iteration round. 
The red line illustrates the accuracy of the CoCoEvo test case population,
the blue line corresponds to the accuracy of its offspring during evolution,
and the gray dashed line reflects the accuracy achieved by the random test case generation approach.
}
\label{fig:tests}
\end{figure*}

Fig. \ref{fig:tests} presents the accuracy trends of the test case population and offspring over the evolutionary iterations, as well as the accuracy of the randomly generated test cases.
As shown in the figure, the accuracy of the test case population consistently exceeds that of randomly generated test cases, demonstrating the effectiveness of the alternating selection strategy in the co-evolutionary framework. 
This indicates that selecting test cases based on their Confidence is beneficial for identifying faulty test cases.
Moreover, the strategy for generating additional test cases exhibits a modest upward trend in the accuracy of offspring test cases, which also contributes positively to maintaining the overall accuracy of the test case population.

Interestingly, the peak test accuracy of the test population is observed in the second iteration.
In subsequent iterations, the accuracy declines slightly.
The slight decline can be attributed to the presence of a few incorrect test cases on the Pareto front that exhibit high discriminative capability but low Confidence, thus affecting overall accuracy.
Nevertheless, the Pareto selection approach preserves a substantial number of accurate and highly discriminative test cases, thereby offering greater benefits to the co-evolutionary process.
Consequently, the marginal reduction in accuracy remains within an acceptable range.

\subsection{Comparison of Program Evolution Scheduler}

\begin{table}[!ht]
\caption{
Comparison of pass@1 performance across different crossover rate schedulers.
\label{tab:cmp-p-sche}
}
\centering
\begin{tabular}{llc}
\toprule
Model & Crossover Rate & Pass@1 \\
\midrule
\multirow{2}{*}{GPT-4o-mini}
& Constant & 45.00 \\
& \textbf{Cosine} & \textbf{49.75} \\
\midrule
\multirow{2}{*}{Qwen2.5-Coder-32B}
& Constant & 51.00 \\
& \textbf{Cosine} & \textbf{55.75} \\
\midrule
\multirow{2}{*}{Llama-3.1-70B}
& Constant & 42.75 \\
& \textbf{Cosine} & \textbf{45.00} \\
\midrule
\multirow{2}{*}{DeepSeek-V3}
& Constant & 72.50 \\
& \textbf{Cosine} & \textbf{76.25} \\
\bottomrule
\end{tabular}
\end{table}

\begin{figure*}[!ht]
\centering
\includegraphics[width=6.6in]{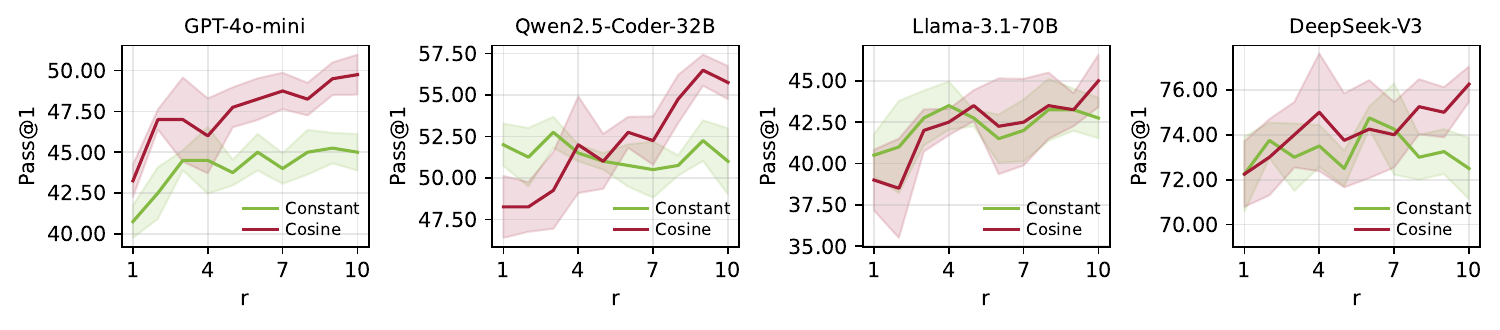}
\caption{
Program pass@1 performance during the evolutionary process for different crossover rate schedulers, where $r$ represents the iteration round.
}
\label{fig:cmp-p-sche}
\end{figure*}

To assess the effectiveness of the crossover rate scheduler, we compare the cosine scheduler with a constant scheduler in our experiments, using a mutation rate of $0.2$ and a crossover rate of $0.8$. 
These experiments were performed on the four LLMs and the LeetCode-Contest dataset.
The results, summarized in Table~\ref{tab:cmp-p-sche} and Figure~\ref{fig:cmp-p-sche}, highlight key differences between the cosine scheduler and the constant rate approach.

At the early stages of the evolution process, the constant method, characterized by a fixed crossover and mutation rate, demonstrates faster convergence and initially outperforms the cosine scheduler on Qwen2.5-Coder-32B, Llama-3.1-70B, and DeepSeek-V3. 
However, this rapid convergence causes the Constant method to stagnate and become trapped in a local optimum, ultimately limiting its overall performance.
In contrast, the cosine scheduler, despite its lower initial performance, maintains consistent improvement throughout the evolution process. 
On GPT-4o-mini, the cosine scheduler consistently outperforms the method with a constant crossover rate.

By adapting the crossover rate dynamically, the scheduler avoids premature convergence and achieves superior results in later stages.
Notably, it surpasses the performance of the Constant method after the fifth iteration, ultimately delivering a higher pass@1 rate.
This comparison underscores the importance of dynamic scheduling in optimizing program evolution, particularly in scenarios prone to local optima.

\subsection{Comparison of Program Fitness Function}
\begin{table}[!ht]
\caption{
Comparison of pass@1 performance across different program fitness functions.
\label{tab:p-fitness}
}
\centering
\begin{tabular}{llc}
\toprule
Model & Program Fitness Function & Pass@1 \\
\midrule
\multirow{2}{*}{GPT-4o-mini} 
& Pass Rate & 45.50 \\
& \textbf{CodeT Score} & \textbf{49.75} \\
\midrule
\multirow{2}{*}{Qwen2.5-Coder-32B} 
& Pass Rate & 53.50 \\
& \textbf{CodeT Score} & \textbf{55.75} \\
\midrule
\multirow{2}{*}{Llama-3.1-70B} 
& Pass Rate & 41.00 \\
& \textbf{CodeT Score} & \textbf{45.00} \\
\midrule
\multirow{2}{*}{DeepSeek-V3} 
& Pass Rate & 73.25 \\
& \textbf{CodeT Score} & \textbf{76.25} \\
\bottomrule
\end{tabular}
\end{table}

\begin{figure*}[!ht]
\centering
\includegraphics[width=6.6in]{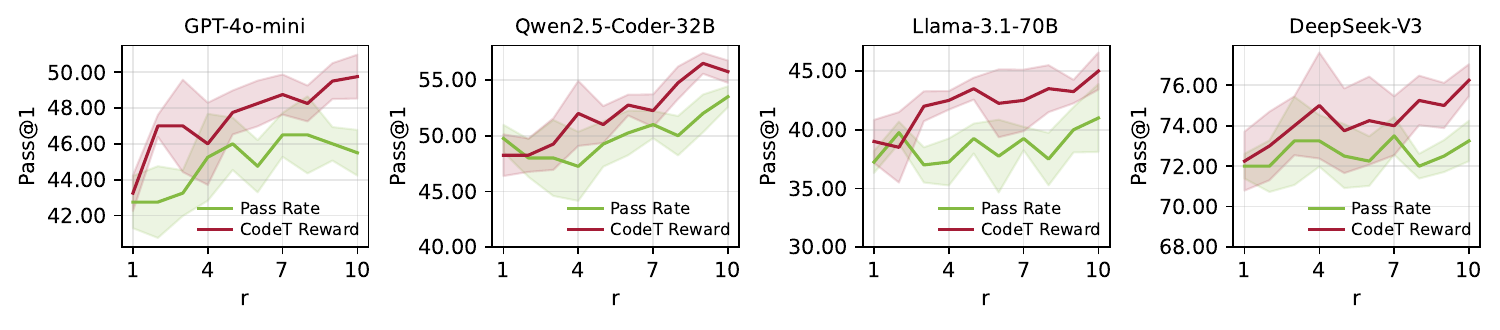}
\caption{
Program pass@1 performance during the evolutionary process for different program fitness functions, while $r$ refers to the iteration round.
}
\label{fig:cmp-p-fitness}
\end{figure*}

To evaluate the impact of different program fitness functions, we conducted a comparative analysis using the CodeT score and the simple pass rate as fitness functions.
These experiments are performed on the four LLMs and the LeetCode-Contest dataset.
The results are summarized in Table~\ref{tab:p-fitness} and illustrated in Figure~\ref{fig:cmp-p-fitness}.

The results indicate that using the raw pass rate as the fitness function yields a lower pass@1 than the CodeT score. 
This is because CodeT better captures the alignment between programs and test cases, reducing the impact of flawed or poorly designed tests on fitness. Consequently, CodeT delivers more stable performance and faster convergence. These findings highlight CodeT’s superiority as a fitness function, particularly when robustness and convergence are essential.

\subsection{Comparison of Test Fitness Function}
\begin{table}[!t]
\caption{
Comparison of pass@1 performance across different test case fitness functions.
\label{tab:t-fitness}
}
\centering
\begin{tabular}{llc}
\toprule
Model & Test Fitness Function & Pass@1 \\
\midrule
\multirow{4}{*}{GPT-4o-mini} 
& Failure Rate & 37.50 \\
& Pass Rate & 47.00 \\
& Confidence & 45.50 \\
& \textbf{Pareto} & \textbf{49.75} \\
\midrule
\multirow{4}{*}{Qwen2.5-Coder-32B} 
& Failure Rate & 37.00 \\
& Pass Rate & 51.50 \\
& Confidence & 51.50 \\
& \textbf{Pareto} & \textbf{55.75} \\
\midrule
\multirow{4}{*}{Llama-3.1-70B} 
& Failure Rate & 26.00 \\
& Pass Rate & 37.50 \\
& Confidence & 36.75 \\
& \textbf{Pareto} & \textbf{45.00} \\
\midrule
\multirow{4}{*}{DeepSeek-V3} 
& Failure Rate & 66.75 \\
& Pass Rate & 71.75 \\
& Confidence & 72.25 \\
& \textbf{Pareto} & \textbf{76.25} \\
\bottomrule
\end{tabular}
\end{table}

\begin{figure*}[!ht]
\centering
\includegraphics[width=6.6in]{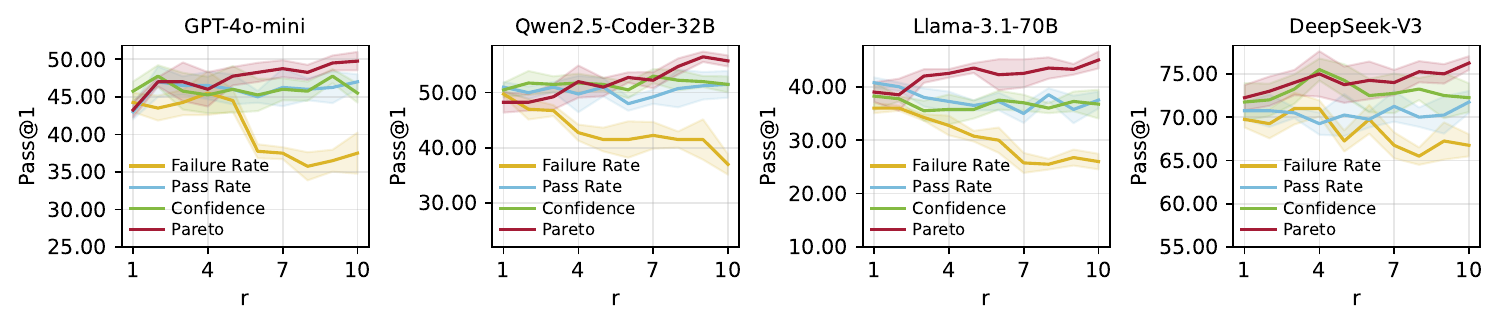}
\caption{
Program pass@1 performance during the evolutionary process for different test case fitness functions, while r refers to the iteration round.
}
\label{fig:cmp-t-fitness}
\end{figure*}

\begin{figure*}[!ht]
\centering
\includegraphics[width=6.6in]{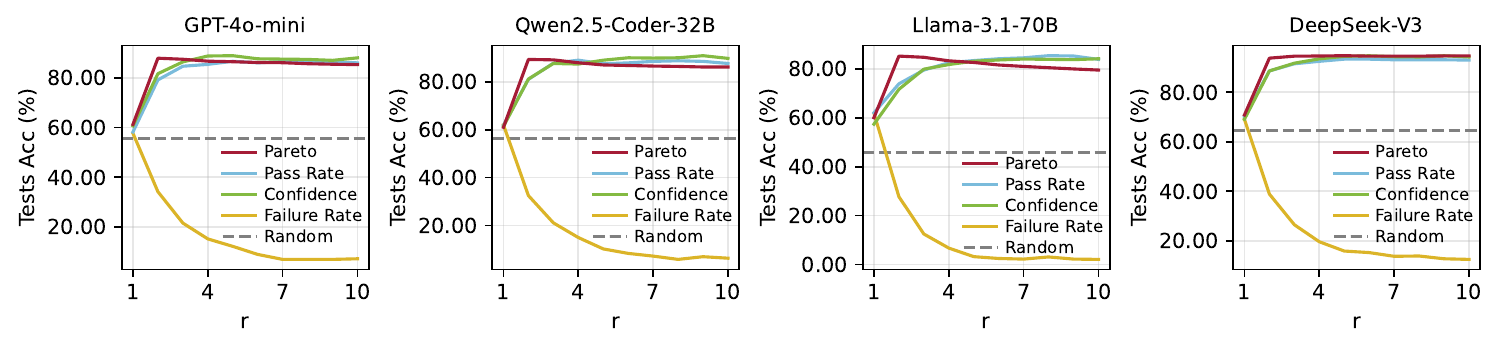}
\caption{
Illustrates the test case accuracy during the evolutionary process for different test case fitness functions, while $r$ refers to the iteration round.
}
\label{fig:cmp-t-acc}
\end{figure*}

In this section, we present the results of comparing various test fitness functions.
Specifically, we evaluate Failure Rate, Pass Rate, Confidence, and Pareto as methods for test fitness calculation.
These experiments are performed on the four LLMs and the LeetCode-Contest dataset. 
The detailed results are depicted in Fig. \ref{fig:cmp-t-fitness}, and additional information on test case accuracy is provided in Fig. \ref{fig:cmp-t-acc}.

Among the test fitness functions, Pass Rate directly calculates the proportion of test cases successfully passed by a program, while Confidence represents a weighted pass rate, as described in Section \ref{sec:test-evo}.
Similarly, Failure Rate measures the failure rate of a program on the test cases.
For these three fitness calculation methods, greedy selection is employed as the offspring selection method.

Experimental results indicate that employing the Pareto method as the selection mechanism yields the best performance. 
In contrast, using Failure Rate as the fitness function results in a sharp decline in pass@1 performance. 
Previous studies on co-evolution have generally argued that the strength of a test case increases as more programs fail on it. 
However, we discovered that if the test cases contain errors, this method can lead to a population dominated by faulty test cases, which fail to provide accurate feedback to the programs.

Furthermore, Fig. \ref{fig:cmp-t-fitness} reveals that employing Confidence as the test fitness function slightly outperforms using Pass Rate in most cases. 
By weighting test cases according to program performance, higher-performing programs contribute more to the Confidence score, ensuring that the most reliable programs have the greatest influence on evaluation.

\subsection{Analysis of Test Coverage Impact}
\label{subsec:test-cov}

\begin{table*}[!ht]
\caption{
Illustrates the impact of test case coverage information on the pass@1 performance of baseline methods and CoCoEvo. 
The suffix ``w/ cov'' in baseline method names indicates the inclusion of test coverage information, while ``CoCoEvo w/o cov'' represents CoCoEvo without test coverage information.
\label{tab:test-cov}
}
\centering
\begin{tabular}{lcccccc}
\toprule
Method & GPT-4o-mini & Qwen2.5-Coder-32B & Llama-3.1-70B & DeepSeek-V3 \\
\midrule
Sampling+Filtering
& 37.50 & 44.50 & 31.00 & 68.25  \\
Sampling+Filtering w/ cov
& 36.25 & 47.75 & 33.25 & 68.50 \\
\midrule
MBR-Exec
& 33.75 & 45.00 & 38.75 & 70.00 \\
MBR-Exec w/ cov
& 35.00 & 45.00 & 33.75 & 72.50 \\
\midrule
CodeT
& 46.25 & 47.50 & 41.25 & 72.50  \\
CodeT w/ cov
& 47.75 & 50.25 & 39.00 & 71.50 \\
\midrule
CoCoEvo w/o cov
& 48.25 & 53.00 & 42.75 & 76.00 \\
\textbf{CoCoEvo}
& \textbf{49.75} & \textbf{55.75} & \textbf{45.00} & \textbf{76.25} \\
\bottomrule
\end{tabular}
\end{table*}

During offspring test case generation, we incorporate coverage information of the current test case population on the best-performing program individual to guide the generation of more targeted test cases.
To explore whether this strategy could improve baseline performance, we applied it to several compatible baselines.
Since this strategy requires an initial set of programs and then iteratively generates test cases based on their coverage,
we adopt this method for Sampling, Sampling+Filtering \cite{li2022competition}, MBR-Exec \cite{shi2022natural}, and CodeT \cite{chen2022codet}.
We also conducted experiments to evaluate the performance of CoCoEvo without the coverage information.
The results are shown in Table \ref{tab:test-cov}.

The experiments reveal that simply adding test coverage information to baseline methods can sometimes negatively affect performance, as observed in GPT-4o-mini with Sampling+Filtering, Llama-3.1-70B with MBR-Exec and CodeT, and DeepSeek-V3 with CodeT. 
In contrast, CoCoEvo benefited consistently from coverage information across all LLMs, underscoring that such data is most effective when used in conjunction with the co-evolutionary framework.

\subsection{Ablation Study}
\label{subsec:ablation-study}

\begin{table}[!ht]
\caption{
Illustrates the results of the ablation experiments.
Where ``w/o test evolution'' represents the method of removing the test case evolution module of CoCoEvo.
\label{tab:ablation}
}
\centering
\begin{tabular}{llc}
\toprule
Model & Method & Pass@1 \\
\midrule
\multirow{3}{*}{GPT-4o-mini} 
& Sampling & 35.25 \\
& w/o test evolution & 48.00 \\
& \textbf{CoCoEvo} & \textbf{49.75} \\
\midrule
\multirow{3}{*}{Qwen2.5-Coder-32B} 
& Sampling & 44.00 \\
& w/o test evolution & 51.75 \\
& \textbf{CoCoEvo} & \textbf{55.75} \\
\midrule
\multirow{3}{*}{Llama-3.1-70B} 
& Sampling & 32.00 \\
& w/o test evolution & 43.00 \\
& \textbf{CoCoEvo} & \textbf{45.00} \\
\midrule
\multirow{3}{*}{DeepSeek-V3} 
& Sampling & 69.00 \\
& w/o test evolution & 74.50 \\
& \textbf{CoCoEvo} & \textbf{76.25} \\
\bottomrule
\end{tabular}
\end{table}

\begin{figure*}[!ht]
\centering
\includegraphics[width=6.6in]{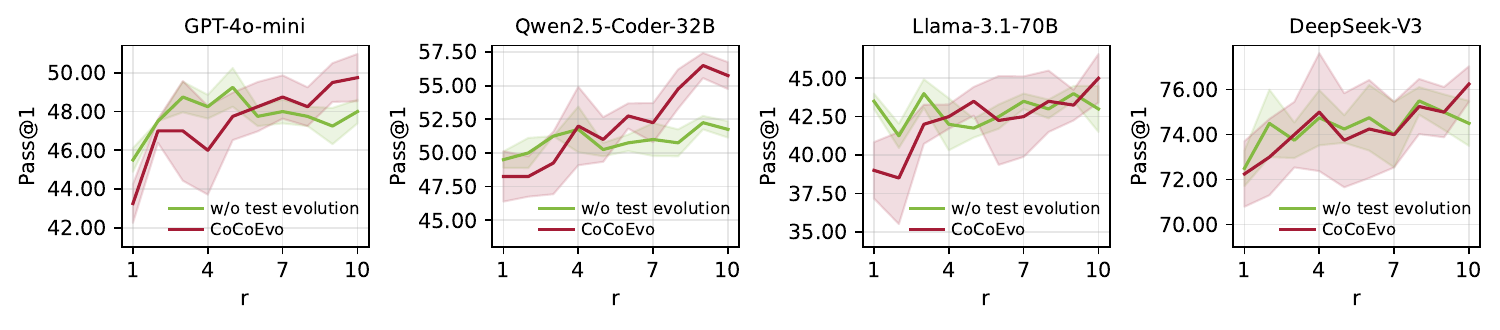}
\caption{
Illustrates the evolutionary process in the ablation study.
The variable $r$ refers to the iteration round.
The red line shows the performance of CoCoEvo, whereas the green line represents CoCoEvo with the test case evolution module removed.
}
\label{fig:ablation}
\end{figure*}

We conduct ablation experiments across the four LLMs on the LeetCode-Contest dataset to evaluate the effectiveness of the test case evolution module. 
In these experiments, we removed the test case evolution module and directly utilized the test cases generated by the LLMs to evaluate the code. 
The comparison results are presented in Table \ref{tab:ablation} and Fig. \ref{fig:ablation}.

The experimental results demonstrate that using only the program evolution module, without incorporating the test case evolution module, still outperforms the Sampling baseline. 
However, after integrating the test case evolution module, the performance of CoCoEvo is further enhanced. 
This highlights the critical role of the test case evolution module in the co-evolution process.

Through the co-evolution of test cases, two key benefits are achieved.
First, the accuracy of test cases within the population improves, as they are refined based on the agreement among programs.
Second, by applying the Pareto method, discriminative and challenging test cases are preserved.
These refined test cases provide more accurate and meaningful feedback to the program evolution process, ultimately improving the program’s pass rate in the final results.

\subsection{Analysis of Token Usage}
\label{subsec:analysis-of-token-usage}

\begin{table*}[!htpb]
\caption{
Illustrates the token consumption across the approaches. 
The average number of input tokens (Prompt) and output tokens (Completion) required for each problem on the LeetCode-Contest dataset are reported. 
The ``Stop-n'' suffix indicates that if, for $n$ consecutive generations, the program population maintains the same fitness and passes the identical set of test cases, the evolutionary process is terminated.
\label{tab:tokens}
}
\centering
\resizebox{\textwidth}{!}{
\begin{tabular}{lcc c cc c cc c cc}
\toprule
\multirow{2}{*}{Method} & \multicolumn{2}{c}{GPT-4o-mini} && \multicolumn{2}{c}{Qwen2.5-Coder-32B} && \multicolumn{2}{c}{Llama-3.1-70B} && \multicolumn{2}{c}{DeepSeek-V3} \\
\cmidrule{2-3}
\cmidrule{5-6}
\cmidrule{8-9}
\cmidrule{11-12}
& Prompt & Completion && Prompt & Completion && Prompt & Completion && Prompt & Completion \\
\midrule
Sampling & 25823.75 & 14326.96 && 26088.75 & 15082.05 && 25900.00 & 13754.51 && 25821.25 & 19535.79 \\
Sampling+Filtering & 28767.25 & 18115.60 && 28932.63 & 18904.11 && 28859.50 & 16305.21 && 28763.38 & 25686.94 \\
\midrule
Self-Repair & 40383.83 & 23086.20 && 40033.05 & 25913.49 && 39661.24 & 23284.41 && 41461.88 & 32701.89 \\
Reflexion & 221277.83 & 48270.31 && 236637.13 & 83716.98 && 234549.55 & 54254.50 && 238286.46 & 57093.93 \\
INTERVENOR & 107996.59 & 79492.75 && 105999.59 & 93253.63 && 131700.33 & 84377.99 && 102779.86 & 85323.98 \\
\midrule
CodeCoT & 44524.69 & 20815.48 && 49635.12 & 19936.05 && 38194.79 & 22017.59 && 41806.95 & 18614.76 \\
AgentCoder & 73915.88 & 35099.35 && 81683.24 & 33976.43 && 71640.85 & 39185.11 && 62136.31 & 27798.86 \\
\midrule
MBR-Exec & 28767.25 & 18115.60 && 28932.63 & 18904.11 && 28859.50 & 16305.21 && 28763.38 & 25686.94 \\
CodeT & 28767.25 & 18115.60 && 28932.63 & 18904.11 && 28859.50 & 16305.21 && 28763.38 & 25686.94 \\
\midrule
CoCoEvo Stop-1 & 13558.04 & 5706.66 && 11414.38 & 5404.08 && 15792.08 & 7535.93 && 10031.14 & 5538.18 \\
CoCoEvo Stop-2 & 20403.58 & 8255.95 && 17998.85 & 8118.11 && 23626.14 & 10699.68 && 15988.85 & 8192.44 \\
CoCoEvo Stop-3 & 26739.18 & 10559.31 && 25055.68 & 11095.44 && 30715.03 & 13666.15 && 22749.20 & 11127.73 \\
CoCoEvo Stop-4 & 33405.58 & 12959.24 && 32192.81 & 14077.55 && 38130.99 & 16786.09 && 30061.74 & 14264.30 \\
CoCoEvo Stop-5 & 40120.51 & 15358.71 && 39038.18 & 16813.46 && 44900.06 & 19619.78 && 37406.18 & 17349.59 \\
CoCoEvo & 61567.61 & 22947.25 && 63638.83 & 26393.06 && 60122.26 & 26128.68 && 66190.20 & 28926.00 \\
\bottomrule
\end{tabular}
}
\end{table*}

\begin{table*}[!htpb]
\caption{
Illustrates the pass@1 performance of the CoCoEvo approach on the LeetCode-Contest dataset under different early stopping strategies. 
The ``Stop-n'' suffix means that if the population’s fitness and passed test set remain unchanged for $n$ consecutive generations, the evolutionary process ends.
\label{tab:tokens-acc}
}
\centering
\begin{tabular}{lcccccc}
\toprule
Method & GPT-4o-mini & Qwen2.5-Coder-32B & Llama-3.1-70B & DeepSeek-V3 \\
\midrule
CoCoEvo Stop-1 & 47.75 & 48.75 & 39.75 & 74.75 \\
CoCoEvo Stop-2 & 47.50 & 51.50 & 39.50 & 76.00 \\
CoCoEvo Stop-3 & 46.00 & 51.00 & 42.75 & 75.25 \\
CoCoEvo Stop-4 & 48.00 & 54.00 & 44.00 & 76.00 \\
CoCoEvo Stop-5 & 48.75 & 54.00 & 43.00 & 75.25 \\
CoCoEvo & \textbf{49.75} & \textbf{55.75} & \textbf{45.00} & \textbf{76.25} \\
\bottomrule
\end{tabular}
\end{table*}

Some strategies, such as reflection and repair employed in some baseline methods, as well as the crossover, mutation, and additional test case generation strategies used in CoCoEvo, may consume more tokens than others. 
We therefore report each method's token consumption to compare their resource efficiency. 
Table \ref{tab:tokens} shows the average token consumption per problem on the LeetCode-Contest dataset. 
We report input (Prompt) and output (Completion) tokens separately, as output tokens are generally more costly. 
Notably, we designed an early stopping strategy to investigate whether the token consumption of the CoCoEvo method could be optimized. 
Since problem difficulty varies, simpler problems often require fewer iterations, 
whereas more complex problems tend to require more. 
Thus, the early stopping mechanism is triggered when the code population maintains the same fitness value and passes the same test cases for $n$ consecutive generations, indicating convergence. 
In Table \ref{tab:tokens}, we report token consumption for different values of $n$ (ranging from $1$ to $5$), denoted as CoCoEvo Stop-1 to CoCoEvo Stop-5, respectively.

As shown in Table \ref{tab:tokens}, Sampling consumes the fewest tokens because it only generates code, not test cases. 
Meanwhile, Sampling+Filtering \cite{li2022competition}, MBR-Exec \cite{shi2022natural}, and CodeT \cite{chen2022codet} show identical token consumption as they share the same generated programs and test cases, differing only in their final program selection logic. 
Code repair methods (e.g., Self-Repair \cite{olausson2023self}, Reflexion \cite{shinn2024reflexion}, INTERVENOR \cite{wang2023intervenor}, CodeCoT \cite{chen2022codet}, and AgentCoder \cite{huang2023agentcoder}) consume significantly more tokens because their prompts must include the original code and test execution results. 
Reflexion consumes the most input tokens because it first generates a reflective analysis before rewriting the code. 
CoCoEvo's input token usage is higher than CodeT's because its crossover, mutation, and test generation modules require existing programs and test cases as input prompts. 
However, its output token usage is lower than that of repair-based methods.

Notably, we observe that with the integration of the early stopping strategy, the token consumption of CoCoEvo is significantly reduced. 
The required input tokens become comparable to those of CodeT, while the output token usage is lower. 
To examine the impact of the early stopping strategy on accuracy, we report the pass@1 performance of CoCoEvo on the LeetCode-Contest dataset under different values of $n$. 
The results are presented in Table \ref{tab:tokens-acc}.
We find that, when jointly considering token consumption and accuracy, setting the early stopping parameter $n$ to $4$ leads to a slight decrease in accuracy compared to configurations without early stopping, but achieves a 30\% to 50\% reduction in overall token usage, demonstrating the considerable potential of CoCoEvo.

\subsection{Evaluation on Real-World Projects}

\begin{table}[!htpb]
\caption{
Illustrates the number of the selected problems and the real-world software projects from which they were drawn.
\label{tab:real-world-dataset}
}
\centering
\resizebox{0.48\textwidth}{!}{
\begin{tabular}{llc}
\toprule
Project & Link & Problems \\
\midrule
mrjob & https://github.com/Yelp/mrjob & 1 \\
Jinja2 & https://github.com/pallets/jinja & 1 \\
twilio-fatisar & https://github.com/twilio/twilio-python & 1 \\
feedparser & https://github.com/kurtmckee/feedparser & 1 \\
dash & https://github.com/plotly/dash & 2 \\
diffprivlib & https://github.com/IBM/differential-privacy-library & 3 \\
databases & https://github.com/encode/databases & 1 \\
\bottomrule
\end{tabular}
}
\end{table}

This section evaluates CoCoEvo's performance on real-world software projects. 
We selected problems from the DevEval dataset \cite{li2024deveval}, where each problem consists of a function and its corresponding unit tests extracted from a real-world project. 
To ensure realistic and manageable problems, we filtered for functions with: (1) at least one external dependency, (2) five or more unit tests, and (3) $10$ to $100$ lines of code. 
In total, we selected $10$ problems from $7$ software projects, as summarized in Table \ref{tab:real-world-dataset}.

\begin{table}[!htpb]
\caption{
Illustrates the comparison of pass@1 between baselines and CoCoEvo on the selected real-world project programming problems.
\label{tab:real-world-acc}
}
\centering
\begin{tabular}{lc}
\toprule
Method & Pass@1 \\
\midrule
Sampling & 38.00 \\
Sampling+Filtering & 40.00 \\
CodeT & 40.00 \\
CoCoEvo & \textbf{44.00} \\
\bottomrule
\end{tabular}
\end{table}

Unlike LeetCode-Contest, these real-world problems have complex dependencies and require full unit tests, not just single assertions. 
To address this, we manually added dependencies to the input prompts, adapted the test generation process to produce full unit tests, and refined the problem descriptions for clarity. 
Using the same setup as Section \ref{subsec:experimental-settings}, we ran experiments with Qwen2.5-Coder-32B \cite{hui2024qwen2} and compared CoCoEvo against three baselines: Sampling, Sampling+Filtering \cite{li2022competition}, and CodeT \cite{chen2022codet}. 
As shown in Table \ref{tab:real-world-acc}, while all methods' performance degraded on real-world problems, CoCoEvo still consistently outperformed the baselines. 
A case study in the supplementary material presents a problem solved only by CoCoEvo, where every baseline fails.

However, our method still exhibits some limitations:
\begin{itemize}
\item Dependency Extraction: Dependencies must be manually supplied as the model cannot automatically identify external modules. Future work could automate this with dependency analysis, LLM agents, or RAG.
\item Context Limitations: Numerous dependencies can exceed model token limits, requiring memory management or larger context windows. 
\item Quality of Problem Descriptions: Real-world projects often have unclear descriptions, challenging the LLM's intent-understanding capabilities. 
\end{itemize}

\section{Conclusion}
\label{sec:conclusion}
This paper introduces CoCoEvo, an LLM-based framework that simultaneously evolves programs and test cases for automated code generation. 
By eliminating the reliance on pre-defined test cases, CoCoEvo addresses a critical limitation in existing approaches, particularly in scenarios lacking comprehensive testing resources.
The dynamic crossover rate scheduling mechanism and the multi-objective optimization for test case selection are key to the framework's success, enabling effective exploration and refinement throughout the evolution process. 
Experimental results on the LeetCode-Contest dataset, utilizing four leading LLMs, demonstrate that CoCoEvo achieves superior performance compared to traditional methods.
Beyond its practical implications in software development, CoCoEvo opens new avenues for integrating evolutionary algorithms with LLMs.

The limitations of CoCoEvo are primarily reflected in its relatively high token consumption. 
However, this can be significantly mitigated by early stopping strategies, which reduce output tokens, though input prompts may remain long. 
Future work could explore how to compress the length of input prompts. 
Moreover, CoCoEvo faces several challenges when applied to real-world software programming tasks. 
Specifically, it lacks precise dependency extraction and memory management capabilities, and it relies heavily on detailed problem descriptions and docstrings. 
Future research may focus on integrating CoCoEvo with techniques such as dependency analysis, LLM-based agents, and RAG, as well as incorporating effective memory management mechanisms, to better adapt the method for real-world software development scenarios. 
In addition, all the datasets used in the experiments of this paper are based on the Python programming language.
However, CoCoEvo is designed language‐agnostic.
Future research may investigate the effectiveness of CoCoEvo across multiple programming languages. 
Furthermore, the evolutionary operators in CoCoEvo rely to some extent on the programming capabilities of LLMs. 
Subsequent work could explore how to adapt CoCoEvo for use with smaller or less capable code language models.

\bibliographystyle{IEEEtran}
\bibliography{paper}

\vfill

\end{document}